\documentclass[aps,prd, twocolumn,nofootinbib, amsmath,amssymb,
preprintnumbers]{revtex4-1}
\usepackage{bm}
\usepackage{hyperref}
\usepackage{amsfonts,graphicx}

\newcommand{\beq}{\begin{eqnarray}}
\newcommand{\eeq}{\end{eqnarray}}
\usepackage{amsmath}
\usepackage{amssymb}
\usepackage[paper=letterpaper,margin=1in]{geometry}
\usepackage{braket}
\usepackage{upgreek}
\usepackage{soul} 
\newcommand{\COMMENT}[1]{\textcolor{cyan}{[#1]}}

\newcommand{\n}{\mathbf n}

\def\det{{\rm{det}}}

\usepackage{tikz}

\begin{document}

\title{Classification of Non-local Actions: Area versus Volume Entanglement Entropy}
\author{Bora Basa$^1$, Gabriele La Nave$^2$, and Philip W. Phillips$^1$}

\affiliation{$^1$Department of Physics and Institute for Condensed Matter Theory,
University of Illinois
1110 W. Green Street, Urbana, IL 61801, U.S.A.}
\affiliation{$^2$Department of Mathematics, University of Illinois, Urbana, Il. 61801}

\begin{abstract}

From the partition function for two classes of classically non-local actions containing the fractional Laplacian,  we show that as long as there exists a suitable (non-local) Hilbert-space transform  the  underlying action can be mapped onto  a purely local theory.  In all such cases the partition function is equivalent to that of a local theory and an area law for the entanglement entropy obtains.   When such a reduction fails,  the entanglement entropy deviates strongly from an area law and can in some cases scale as the volume.  As these two criteria are coincident, we conjecture that they are equivalent and provide the ultimate test for locality of Gaussian theories rather than a simple inspection of the explicit operator content. 

\end{abstract}

\maketitle
\section{Introduction}
Locality of the action is a fundamental tenet of quantum and effective field theory.  In fact, the well known area law\cite{bombelli,Srednicki,heatkernel,callanGeometricEntropy1994,Hawking2001,Calabrese,rt2006,ch2009,eisert} for the entanglement entropy (EE) is a direct consequence of the locality (near-neighbour interactions) of the action.  Deviations then from the area law are expected to obtain as non-local interactions are introduced.  However, this problem is quite subtle as the work of Li and Takayanagi\cite{lt2011,st2014} demonstrates.  They considered two non-local actions, B) $I_B(\phi)= \int\,  d^d x \,\phi (-\Delta)^\gamma \phi$ and  C) $I_C(\phi)= \int\,  d^d x \, \phi e^{ (-\Delta)^\gamma}\phi $ with $\gamma\in {\mathbb R}$, where $(-\Delta)^\gamma$ is the fractional Laplacian.  Although both of these theories contain non-local operators, they display fundamentally different scaling of the zero-temperature entanglement entropy:
\beq\label{entropy}
    S_B&\sim& \kappa_{d-2}\left(\frac{1}{\epsilon}\right)^{d-2}+\cdots,\quad{\rm B-theories}\nonumber\\
    S_C&\sim&\kappa_{d-2} \left(\frac{1}{\epsilon}\right)^{d-2+2\gamma}+\cdots\quad{\rm C-theories}
\eeq
where $\epsilon$ is a short-distance cut-off and $\kappa_{d-2}$ is an arbitrary function defined on the entangling surface that doesn't affect the scaling argument.  As is evident, in $B$-type theories, the EE has the typical area scaling of a local QFT,  understood as the entropic contribution of UV degrees of freedom that are entangled across a separating surface, $\Sigma$ on the Cauchy slice \cite{bombelli,Srednicki,heatkernel,callanGeometricEntropy1994,Hawking2001,Calabrese,rt2006,ch2009,eisert}.  However,  $C$-type theories (see also\cite{levine}) deviate strongly from this scaling and can in the case of $\gamma=1/2$ yield a volume law.  Hence, not all non-localities in the action give rise to deviations from area laws. Precisely what is the criterion for the transition between these types of theories or the conditions for a change from area to volume EE has never been clarified.  This problem is also relevant to neutron-star collapse as a transition has been observed\cite{alexander2018} between volume and area laws for the EE.

This paper lays plain the precise types of non-localities that preserve the area law.  We find that the minimum requirement for turning B-type into C-type theories is the introduction of a (fractional) mass term, hence a minimal action of the form $I(\phi)= \int\,  d^d x \,(\phi (-\Delta)^\gamma\phi+m^2 \phi^2)$.  In the absence of the mass, it is possible to recast all B-type theories via a Hilbert-space transformation as purely local theories.  The exponential in $I_C$ is just an extreme case of this non-reduction.  At the level of the EE, we can think of the mass term as providing a necessary scale for probing the fractional structure of the entangling surface. In the absence of such a scale, the UV physics remains insensitive to the choice of $\gamma$. As a result, we conjecture that these two criteria, the presence of a local Hilbert-space transformation and the area law are equivalent and ultimately determine whether the action for a QFT is truly local.

Although non-localities typically indicate that  something went terribly wrong\cite{hertz,millis,sungsik,belitz}, for example the non-Wilsonian procedure of integrating out gapless degrees of freedom, they are oftentimes  fundamental. In fact, a non-perturbative theory of quantum gravity is unlikely to be defined through the usual local structures. Rather, locality likely emerges from a broader organizing principle that doesn't demand it \emph{a priori}. The Caffarelli-Silvestre (CS) extension theorem\cite{CS2007} demonstrates that second-order elliptic differential equations in the upper half-plane in ${\mathbb R}_+^{n+1}$ reduce to one with the fractional Laplacian, $(-\Delta)^\gamma$ at  ${\mathbb R}^n$, where a Dirichlet boundary condition is imposed.  Quite generally, the fractional Laplacian $(-\Delta)^\gamma$ (or its conformal extension, the Panietz operator\cite{graham,gz,graham}) on a function $f$ in ${\mathbb R}^n$ provides a  Dirichlet-to-Neumann map for a function $\phi$ in ${\mathbb R^{n+1}}$ that satisfies the second-order elliptic differential equation. In our notation, the fractional Laplacian in the absence of a mass term should be understood as the conformal Laplacian raised to a non-integer power. All the non-localities we consider will be constructed from the fractional Laplacian which has numerous uses in holography\cite{gz,lnp1,lnp2,rajabpour,gonzalez,Gross2017} and long-ranged statistical models\cite{slava,Blanchard}. Note also that Lagrangians with kinetic terms of the form $\phi \eta(\Delta)\phi$ may be non-local in general for certain choices of the function $\eta$ even if the Laplacian is not raised to a fractional power. We do not consider such theories here. 
\section{Path integral quantization}
\subsection{Lattice Integral}

To illustrate our main point in the most familiar of settings, we consider the two Gaussian integrals,
\beq Z=\int _{\mathbb R ^d} \, dx_1\cdots dx_d\; e^{-\frac{1}{2} x\left(M+m^2\bf 1\right)^\gamma x +Jx}\eeq
and
\beq Z_\gamma=\int _{\mathbb R ^d} \, dx_1\cdots dx_d\; e^{-\left(\frac{1}{2} xM^\gamma x +m^2 x^2\right)+Jx}\eeq
where of course $x^2=x\cdot x$ and $M$ is some diagonalizable matrix.
Formally,  $Z$ and $Z_\gamma$ are the very similar in that they can both be dealt with by replacing $(M+m^2 \bf 1)^\gamma$ and $M^\gamma +m^2 \bf 1$ by some matrtix $U$ and, then solving the Gaussian integral by diagonalizing $U$ and finally analyzing all the various eigenvalues.
However, there is a major difference between these two path integrals.  The transformation
\beq x&\to& \left(M+m^2\bf 1\right) ^{\frac{1-\gamma}{2}}x\nonumber\\
J&\to& \left(M+m^2\bf 1\right) ^{\frac{1-\gamma}{2}}J
\eeq
maps $Z$ onto a purely local Gaussian theory up to a constant that depends only on $\det (M+m^2\bf 1)$.  As we will see, no such field redefinition  which effectively removes the non-locality is possible for $Z_\gamma$.  This effective Hilbert-space transformation plays out in the field quantization and
the computation of the EE.

Despite this difference, both of these theories can be quantized.
We intend to show that
\beq Z= \left( \frac{(2\pi)^d}{\det (M+m^2 \bf 1)^\gamma}\right)^{\frac{1}{2}}\, e^{\frac{1}{2} J\cdot \left(M+m^2\bf 1\right)^{-\gamma}\cdot J}\eeq
and
\beq Z_\gamma= \left( \frac{(2\pi)^d}{\det (M^\gamma+m^2\bf 1)}\right)^{\frac{1}{2}}\, e^{\frac{1}{2} J\cdot \left(M^{\gamma}+m^2\bf 1\right)^{-1}\cdot J}.
\eeq
To proceed, we note that $M^\gamma$ and $M$ commute, and therefore $[M,\left(M+m^2\bf 1\right)^\gamma]=0$ and $[M, M^\gamma+m^2 {\bf 1}]=0$. Thus, they
can be simultaneously diagonalized (along with $\left(M+m^2\bf 1\right)^\gamma$ and $ M^\gamma+m^2 \bf 1$) and therefore it is possible to find an orthogonal matrix $O\in O(n)$ such that $M= O^{-1}D\, O$,
where $D$ is a diagonal matrix, $D_{ij}= \lambda _i \delta _{ij}$ and that
$M^\gamma = O^{-1}D^\gamma\,O$.  With this in hand, we find that $ \left(M+m^2\bf 1\right)^\gamma=O^{-1}\left(D+m^2\bf 1\right)^\gamma\,O$ and  $M^\gamma+m^2{\bf 1}=O^{-1}\left(D^\gamma+m^2 \bf 1\right)\,O$. 

With this result, we can then perform the integrals explicitly by
changing coordinates.  To this end, we define
 $y^\intercal= O x^\intercal$, where $y= (y_1,\cdots , y_d)$ and $x= (x_1,\cdots , x_n)$. After this change of coordinates, the integral transforms to
\beq\label{local-descreteGauss}
Z= \int _{\mathbb R ^d} \, dy_1\cdots dy_d\, \det (O)\; e^{-\frac{1}{2} yD^\gamma y +JO^{-1}y}\nonumber\\
\int _{\mathbb R ^d} \, dy_1\cdots dy_d \;e^{ -\sum _{\ell=1}^d \lambda _\ell ^\gamma y_\ell^2+\sum _{\ell=1}^d \, j'_\ell y_\ell}\nonumber\\
 =\prod _{\ell=1}^d \left( \left(  \frac{2\pi }{\lambda _\ell^\gamma}  \right)^{\frac{1}{2}} \, e^{-\frac{{j'_\ell}^2}{2\lambda _\ell^\gamma} }\right)\\= \left(  \frac{(2\pi )^{\frac{d}{2}}}{\det(M^\gamma)^{\frac{1}{2}}}  \right) e^{\frac{1}{2} J\cdot M^{-\gamma}\cdot J},
 \eeq
where we have set $J'= JO^{-1}= J O ^\intercal$ and have used that $O\in O(n)$, and thus $O^{-1}=  O ^\intercal$ and $\det (O)=1$. The same calculation can be tailored to the second formula.
Proceeding, we obtain
\beq
\label{nonlocal-descreteGauss}
Z_\gamma= \int _{\mathbb R ^d} \, dy_1\cdots dy_d\, \det (O)\; e^{-\frac{1}{2} yD^\gamma y+m^2y^2 +JO^{-1}y},\nonumber\\
\int _{\mathbb R ^d} \, dy_1\cdots dy_d \;e^{ -\sum _{\ell=1}^d \left(\lambda _\ell ^\gamma+m^2\right) y_\ell^2+\sum _{\ell=1}^d \, j'_\ell y_\ell}\nonumber\\ =\prod _{\ell=1}^d \left( \left(  \frac{2\pi }{\left(\lambda _\ell ^\gamma+m^2\right)}  \right)^{\frac{1}{2}} \, e^{-\frac{{j'_\ell}^2}{2\left(\lambda _\ell ^\gamma+m^2\right)} }\right)\nonumber\\=\left( \frac{(2\pi)^d}{\det (M^\gamma+m^2\bf 1)}\right)^{\frac{1}{2}}\, e^{\frac{1}{2} J\cdot \left(M^{\gamma}+m^2\bf 1\right)^{-1}\cdot J}.\nonumber\\
\eeq

The true non-locality of the second theory, i.e. $Z_\gamma$, is manifest when one tries to compare it with the known local theory (or rather the one known to be equivalent to a local theory by what we just proved). In doing this, one has to analyse the expression
\beq
e^{\frac{1}{2} J\cdot \left(M^{\gamma}+m^2\bf 1\right)^{-1}\cdot J}.
\eeq
In fact one finds, by simple algebra, that, so long as $\Vert m^2M^{-\gamma}\Vert <1$, that is to say so long as $\Vert M^\gamma\Vert >m^2$
\begin{footnote}{There is a similar expansion for $\Vert M^\gamma\Vert <m^2$, but $\Vert M^\gamma\Vert >m^2$is true away from a finite dimensional vector subspace of the full Hilbert space.}
\end{footnote}
\begin{equation}\label{expansion} \frac{1}{2} J\cdot \left(M^{\gamma}+m^2\bf 1\right)^{-1}\cdot J=\frac{1}{2} \sum _{k=0}^{\infty} m^{2k} JM^{-\gamma k}J\end{equation}
thus giving rise to an infinite tower of ``local'' theories (hence the non-locality, which is akin to the structure of the fractional Virasoro algebra of \cite{lnp4}. The same is also true for the Lagrangian involving $e^{-\Delta ^\gamma}$ considered by Li and Takayanagi \cite{lt2011}.

\subsection{Functional field integral}
The previous analysis enables an immediate quantization of the underlying field theories.
We consider the partition function
\beq
Z[J]= \int \, \mathcal D\phi \, e^{i\int\, d^d x\, [\frac{1}{2} \phi (-\Delta +m^2)^\gamma \phi+ J\phi]}.
\eeq
The field redefinition  $\psi= (-\Delta ^2+m^2)^{\frac{1-\gamma}{2}} \phi$ followed with $J'= (-\Delta ^2+m^2)^{\frac{1-\gamma}{2}} J$ maps this action onto a Gaussian model.  To obtain the original action under this transformation, it is necessary to integrate the transformed action by parts.  Such an integration by parts rule exists for any power of the Laplacian but  an exponential of the fractional Laplacian would require infinitely many Hilbert-space transforms (see Eq. (\ref{expansion})) thereby making any transform to a local theory impossible.   As a result, no such field redefinition exists for C-type theories. As a concrete example, consider a transformation that localizes the kinetic term of the action
\begin{equation}
    I = \int_{\mathbb{R}^d}d^dx \phi( (-\Delta)^\gamma +m^2)\phi + J\phi.
    \label{eq:typeC}
\end{equation}
Under the field redefinition, $\psi = (-\Delta)^{\frac{1-\gamma}{2}} \phi$ and the corresponding transformation of the current, the new action becomes 
\begin{equation}
    I' = \int_{\mathbb{R}^d}d^dx \psi \left(-\Delta+\frac{m^2}{ (-\Delta)^{1-\gamma}}\right)\psi + J'\psi.
\end{equation}
As seen clearly, localizing the kinetic term dynamically antilocalizes the quadratic self-interaction. One could of course argue that this was a n\"aive expectation and that one should require that $\int_{\mathbb{R}^d}d^dx \phi( (-\Delta)^\gamma +m^2)\phi = \int P\phi P\phi$ for some pseudodifferential operator $P$ (identity to which we refer as {\it integration by parts}),  which would necessarily have to have order $\frac{\gamma}{2}$. In fact, one can convince oneself rather quickly that this is impossible, by writing $P= P'+Q$ with $P'$ arising from the symbol of $P$ and with $Q$ of smaller order. This is of course the rather obvious fact that a theory that has a kinetic energy term and "potential" term cannot be written as a theory that only has a generalized kinetic energy term.

A sufficient rule for the existence of a localizing Hilbert-space transformations is
\beq\label{rule}
\rm spec\{\hat O(\gamma)\}=\rm spec\{\hat O\}^\gamma,
\eeq
where $\rm spec$ stands for the spectrum of the eigenvalues, assuming it is discrete, and $\hat O(\gamma)$ is the generalized kinetic energy operator.  In case $\hat O$ has a discrete spectrum (e.g., self-adjoint on a compact manifold), this is equivalent to requiring that  $\hat O (\gamma)= \hat O^\gamma$ where $\hat O^\gamma= \frac{1}{\Gamma(-\gamma)}\int_0^{+\infty} \frac{dt}{t^{1+\gamma}}\, e^{-t\hat O}$, where $e^{-t\hat O}$ is the diffusion semigroup associated to $\hat O$. Eq. (\ref{rule}) clearly fails for all C-type theories.  Note the exponential kinetic term in C-type theories violates Eq. (\ref{rule}) even for $\gamma=1$ in which only the Laplacian is present in the exponent.   In addition, the $\cos\partial_\mu$ kinetic term used by Levine\cite{levine} (which generates volume EE) violates  Eq. (\ref{rule}) thereby lending further evidence that Eq. (\ref{rule}) must hold for the area law to obtain.  Consequently, we propose that an action $I$ is {\it local} if its path integral is equivalent (i.e. equal up to a constant) to a classically local action. An elementary calculation (following the calculations we did for $\hat O=\Delta$) should convince the reader that this holds for $S(\phi) =\int \; \phi \hat O(\gamma)\phi$ for an operator $\hat O(\gamma)$ such that $\hat O(\gamma)=\hat O^\gamma$ with $\hat O$ classically local.

To compute the path integral we will need the fractional propagator,
\beq \label{eq:frac_prop}
(-\Delta ^2 +m^2)^\gamma D_\gamma(x-y)= \delta ^d(x-y),
\eeq
The path integral will involve the determinant of such an operator.  This will be evaluated using the standard $\zeta$-function regularization procedure~\cite{seeley,hawking} .
  Let $M$ be an elliptic, self-adjoint operator, so that it has a complete spectrum. Let $\{ \lambda _n\}$ be the sequence of its eigenvalues: $M\phi _n=\lambda _n \phi _n$. The goal is to define $\det(M)$ by $\zeta-$function regularization (essentially following~\cite{seeley}). Given a sequence of eigenvalues $\{ \lambda _n\}$ one can form the (generalized) zeta function:
\beq
\zeta (s)= \sum _n \lambda _n^{-s}.
\eeq
It is a standard fact that $\zeta(s)$ is convergent for $Re(s)>2$ and that it can in fact be extended analytically to a meromorphic function throughout the entire complex plane $\mathbb C$ with poles only at $s=0$ and $s=1$.
Next observe that on the one hand
\beq
\frac{d}{ds} {\rm tr} (M^{-\gamma s} )= \frac{d}{ds} \sum _n \lambda _n^{-\gamma s}=\frac{d}{ds}  \zeta (\gamma s)
\eeq
and that on the other
\beq
\frac{d}{ds} \sum _n \lambda _n^{-\gamma s}=  \sum _n \; (-\gamma \log \lambda _j) \lambda _j^{-\gamma s}
\eeq
whence, formally, for $s=0$
\beq
\frac{d}{ds}\bigg\vert _{s=0} \sum _n \lambda _n^{-\gamma s}= -\sum _n \; \lambda _j^\gamma,
\eeq
which equals (formally) $\log \det (M^\gamma)$.  We thus define
\beq
\det (M^\gamma)= {\rm exp} \left( \frac{d}{ds}\bigg\vert _{s=0} \zeta (\gamma s)\right)
\eeq
which is the $\zeta$-function regularization of $\det (M^\gamma)$. This regularization scheme naturally works for the fractional Laplacian on a curved manifold, giving rise to a generalization of \cite{hawking} to fractional Laplacians.  The path integral is now given by
\beq
\label{freefractint}Z[J] = \frac{1}{\sqrt{\det (-\Delta)^\gamma}}  e^{iW(J)}
\eeq
where
\beq
W(J) = -\frac{1}{2} \int _{\mathbb R^{d}}d^d xd^d y \, J(x) D_\gamma (x-y) J(y)
\eeq
and $D_\gamma(x-y)$ is the {\it fractional} propagator defined in Eq.~(\ref{eq:frac_prop}).

For the C-type theories, the partition function is given instead by
\beq
\label{truefreefractint}Z_\gamma[J] = \frac{1}{\det \left((-\Delta)^\gamma+m^2\right)}  e^{iW_\gamma(J)},
\eeq
where
\beq
W_\gamma(J) = -\frac{1}{2} \int _{\mathbb R^{d}}d^d xd^d y \, J(x) \tilde D_\gamma (x-y) J(y)\nonumber\\
\eeq
and $\tilde D_\gamma(x-y)$ is the {\it fractional} propagator
\beq
\left(-\Delta^\gamma +m^2\right)\tilde D_\gamma(x-y)= \delta ^d(x-y).
\eeq
Armed with these examples we propose the following criterion of non-locality: A QFT is truly non-local if there is no transformation of the Hilbert spaces (even possibly defined away from a finite dimensional vector space) which casts the theory as a finite sum of local theories.  This definition clearly sets type-B and type-C theories apart.

Though we are presently limited to Gaussian theories since we would like to make contact with EE calculations, it is natural to consider arbitrary deformations of the fractional Gaussian fixed point. The apparent locality of the Gaussian fixed point has important implications for the way one interprets the space of theories surrounding it, in particular the non-trivial fixed points. Unlike the Gaussian case that we are presently considering where we can complement the discussion with EE scaling arguments, the subtleties of non-locality in interacting theories may be put on firm footing first by clarifying not only the analytical properties (at the level of OPEs) of the theory and the underlying (non-local) CFT~\cite{confNL} but also the algebraic extension one must make in order to accommodate non-local CFTs (and the operator algebra in general). The former has been thoroughly discussed, particularly in the long-range Ising model literature (Refs~\cite{slava} and certain references therein) while the latter is the subject of Ref.~\cite{lnp4} and future work.

\section{Entanglement entropy}
In this section we determine the leading divergence of the EE for the non-local theory described by $Z_\gamma$.
It is well established that local quantum field theories\cite{rt2006,ch2009} have entanglement entropies that scale as the area of the entangling surface. Though certain features of this scaling law depend on the specifics of the regulators of the theory, quite generally, one has that for a local $d$ dimensional field theory, the leading UV divergence is given by the first of Eqs. (\ref{entropy}).  


\section{Geometric entropy in a QFT and in a CFT}\label{sec:replica}

Let $(M,g)$ be a  globally hyperbolic space-time such that $M$ is diffeomorphic to $\mathbb R \times K$. Let $\partial A=\partial \bar A=\Sigma\subset K\subset M$ be an {\it entangling surface} separating two regions $A$ and $\bar A$ on a Cauchy slice, $K$. We want to calculate the geometric entropy of the QFT,
\beq
Z= \int \; \mathcal {D}\phi e^{-S[\phi]},
\eeq
across $\Sigma$ following Wilczek {\it et al.}~\cite{callanGeometricEntropy1994}. For the sake of simplicity of the argument we assume that the normal bundle $N_{\Sigma/K}$ is trivial (i.e., $\Sigma$ is orientable since it is codimension ~1 ), and therefore so is the normal bundle $N_{\Sigma/M}$ since $M$ is a product manifold.
In this discussion the action, $I$, corresponds to either theories of type B or C and is not assumed to be {\it local} in any sense. We merely require that $I$ preserve the Hilbert space decomposition $H= H_A\hat \otimes H_{\bar A}$ ($\hat \otimes$ reads "completion of the tensor product")\begin{footnote}{This is problematic because no such tensor product decomposition exists from the point of view of algebraic QFT. As is common in  literature, we ignore this technical difficulty by appealing to the fact that the class of theories we consider are free. }\end{footnote} A potential issue regarding the compatibility of the (fractional) regulator with the Cauchy slice bipartition is that it is not immediately obvious that this partition induces a Hilbert space bipartition when the degrees of freedom are non-local. This is resolved by the Caffarelli-Silvestre extension theorem: The non-local theory in question is local in one higher dimension and can be partitioned there, provided we give boundary conditions on the entangling surface extended to this higher dimensional space.
Let $\tau$ be imaginary time, so that $K=\{\tau =0\}$. The vacuum state of the QFT is defined as 
\beq 
\Phi (\phi_0) = \int_{\phi\mid_{\tau=0}= \phi_0} \; \mathcal {D}\phi e^{-I[\phi]},
\eeq
where $\phi_0$ is a field on slice $K=\{\tau=0\}$. We define also $\phi_\pm$ on time slices slightly deformed away from $\tau=0$, $K=\{\tau=0^{\pm}\}$.
Since the normal bundle is trivial we can fix a global unit normal vector $\mathbf n$ which we imagine separating $K$ into two regions depending on whether $\n$ is positive or negative ($A$ and $\bar A$ resp.). 
The choice of $\n$ also gives rise to a Hilbert subspace of $H_K$ (the Hilbert space of $K$). One defines $H_\pm^A $ to be the space of fields which are equal to $\phi_\pm^A$ (the vacuum) on $A$ and analogously $H^{\bar A}_\pm$ the ones that are equal to$\phi_\pm$ on $\bar A$. 
The reduced density matrix for a region $A$ is defined by 
\beq
\rho_A=\frac{1}{Z}\int \mathcal{D}\phi^{\bar A}_0 \Phi^*(\phi^{\bar A}) \Phi(\phi ^{\bar A}),
\eeq
where the boundary conditions are such that the trace over $\phi^{\bar A}$ identifies $\phi^{\bar{A}}_{+}$ and $\phi^{\bar{A}}_{-}$ and acts trivially on the regulated cut. The field can be thought of as being defined on the whole $\mathbb R \times K$ (in Euclidean signature, after a Wick rotation) except on $ A$ (the cut).
In order to consider the trace of $\rho ^n$, which will be of interest in computing the R\'enyi entropy, we need to better understand the geometric construction of such an object. We consider $\mathcal N$ to be the total space of the normal bundle  $N_{\Sigma/M}$ and set $\mathbf \tau$ to be the (unit norm) section of $N_{\Sigma/M}$ corresponding to $\tau$. Then $M$ can be identified as $\n=0$ and $K$ as $\n=T=0$ in $\mathcal N$. We chose a ramified (branched) covering of degree $n$ of $M$ ramified at $\Sigma$. 
This can be thought of as living in $\mathcal N \times \mathbb R$ and there are many non-equivalent such objects which do not affect the calculation. Call such spaces $\mathcal C_n$. 
The metric on $\mathcal C_n$ is conic along $\Sigma$ with conic angle $2\pi (1-n)$.  By the replica trick,
\beq
Tr \rho ^n = Z[\mathcal C_n],
\eeq
where $Z[\mathcal C_n]$ is the Euclidean integral pulled-back to $\mathcal C_n$. In practice, defining well-behaved fields on the covering space involves a diagonalization procedure in the space of possibly non-local replica fields. As in the usual local case, if the theory is quadratic, this poses no concern for the class of theories we consider because $[\Delta, \Delta^\gamma]=0$. We then perform an analytic continuation by choosing any metric $g_\delta$ on $M$ which is conic along $\Sigma$ of conic angle $\delta\in [0,2\pi]$. Such a family of metrics cannot be chosen arbitrarily because, at a minimum, the condition $\lim _{\delta \to 1} g_\delta =g$ must be met to preserve the $n=1$ physics.
Then the geometric entropy is defined as 

\beq
    S_\delta=-(2\pi \partial_\delta +1)\log Z_\delta,
\eeq

If $\delta \to 0$ is a unique analytical continuation, the geometric entropy, as defined, computes the EE,
\beq
S=-\text{Tr}\rho_A\log \rho_A.
\eeq

If the theory in question is a CFT, the \emph{formal} story remains the same: One must compute the replica partition function of a $CFT_{d}$. For $d>2$, we simply do not have the computational machinery to handle such a task except in very specialized cases. Algebraically, the problem lies in the fact that while one may formally envision twist operator insertions that encode the branched structure of replica space,  one cannot carry the analysis further except in specialized cases~\cite{HUNG2014}. The geometric dual of this algebraic complication is that, unlike the two dimensional case, the \emph{general} rammified geometry that results from the replica trick does not admit a closed form partition function~\cite{VASSILEVICH2003279}. As we shall elaborate upon shortly, when the objective is to compute the scaling law of the EE, the finite correlation length ensures that we can compute the replica partition function asymptotically. In particular, the dominant geometric features that appear in the scaling law are related to the cone radius and the curvature. If the theory is scale invariant, such an approach is not possible since the asymptotics cannot be controlled in a physically meaningful way.

\subsection{EE scaling of massive theory}
The effective action, $F$, on $C_\delta\times \Sigma$, is given by a Gaussian path integral: $-\log Z_\delta=\log \det (-\Delta^\gamma+m^2)$. For simplicity, let us assume $M=\mathbb{R}^d$ and $\Sigma=\mathbb{R}^{d-2}.$ To compute the functional determinant in the effective action, we use the heat kernel method\cite{J}\begin{footnote}{For a recent review, we refer the reader to Ref.~\cite{nishioka}wherein the authors walk through the computation of the EE as a $1/m$ expansion of the heat kernel. }\end{footnote} with a hard UV cut-off,
\begin{equation}\label{F-mass}
    F = \int^\infty_{\epsilon^{2\gamma}} \frac{ds}{s} \text{Tr}e^{s \Delta^{\gamma}}e^{-sm^2}.
\end{equation}
  The fractional power of the short distance cut-off is for dimensional consistency.  The heat-kernel method is just as well suited to study non-local theories~\cite{nesterov1,nesterov2} as it is for local ones.
The trace of the heat kernel, $\zeta(s):=\text{Tr}e^{-s \Delta^\gamma }$, factorizes naturally  on the underlying product space.
For the fractional heat kernel on the cone of radius $R\sim m^{-1}$, the asymptotics are (cf. the appendix)
\begin{equation}\label{conicheatexp}
\begin{aligned}\zeta_{C_\delta}(s)&= s^{-1/\gamma}R^2 \frac{2\pi-\delta}{8 \pi}+Cs^{-1/\gamma +1}\\&+\mathcal{O}\left(\frac{s^{1/\gamma}}{R^2}\right) .\end{aligned}
\end{equation}
The constant, $C>0$, is related to the curvature of the cone and does not contribute to the UV scaling of the entropy. On the flat entangling surface with $d-2>0$, it may be computed directly via Fourier transformation,
\begin{equation}
  \begin{aligned}
    \zeta_{\Sigma^{d-2}}(s) &= \frac{A_{d-2}}{\Gamma(d/2-1)2\gamma}\Gamma\left(\frac{d-2}{2\gamma} \right)\int_0^\infty \frac{dp p^{d-3}}{(2\pi)^{d-2}} e^{-sp^{2\gamma}}\\
    &=s^{\frac{2-d}{2\gamma}}\frac{A_{d-2}}{\Gamma(d/2-1)2\gamma(2\pi)^{d-2}}\Gamma\left(\frac{d-2}{2\gamma} \right),
  \end{aligned}
\end{equation}
where $A_{d-2}=\text{Area}(\Sigma)$. If we assume that we can uniquely continue $S_\delta$ to non-integer $\delta$ with $\text{Re}(\delta)>0$ and take the limit $\delta\to 0$\cite{Hardy1920}, the EE becomes
\begin{equation}\label{mass-ent1}
  \begin{aligned}
    S &= \int_{\epsilon^{2\gamma}}^{\infty}\frac{ds}{s}\left[(2\pi \partial_\delta +1)\zeta_{C_\delta}\right]_{\delta\to 0}\zeta_{\Sigma^{d-2}}e^{-s m^2}\\
    &= \kappa_{d-2}\int_{\epsilon^{2\gamma}}^{\infty}\frac{ds}{s} s^{1-\frac{d}{2\gamma}}e^{-s m^2},
  \end{aligned}
\end{equation}
which implicitly requires $m\ne 0$ so that one can define a cone scale. In the final expression, we labeled the multiplicative factors as $\kappa_{d-2}$. We may infer from the form of Eq. \eqref{mass-ent1} that the necessary and sufficient condition for an operator $\mathcal {O}$ to have a heat kernel with an area law is
\beq\label{equiv}
(2\pi \partial_\delta +1) \text{Tr}e^{s \mathcal O(\gamma)}= s^{\frac{2\gamma}{d}} \, \sum _{k=0}^{\infty} a_k s^k,
\eeq
where the heat kernel trace $\text{Tr}e^{s \mathcal O}$ is calculated on $C_\delta\times \Sigma$.

For small $\epsilon$, the leading-order divergence is an area law violation:
\begin{equation}\label{mass-ent2}
  \begin{aligned}
    S &\sim \kappa_{d-2}\Gamma\left(1-\frac{d }{2 \gamma },m^2 \epsilon ^{2 \gamma }\right)\\
    &\sim \kappa_{d-2} \left(\frac{1}{\epsilon}\right)^{d-2\gamma}+\cdots. \\
  \end{aligned}
\end{equation}
Here we kept only the terms that scale with $\epsilon$ and carried out an asymptotic expansion of the incomplete gamma function for small $\epsilon$, corresponding to the UV limit. The volume law appears when $\gamma=1/2$, while the area law for $\gamma=1$, which is the conventional free-theory limit. This flat space calculation can be carried out for any entangling surface in a globally hyperbolic spacetime.

\subsection{EE scaling of massless theory}

The heat kernel expansion from which Eq. \eqref{mass-ent1} follows can be interpreted as a $1/m$ expansion~\cite{nishioka} as the mass scale is the only meaningful scale in the problem. Thus, one should not expect $m\to 0$ to produce the correct EE scaling for the massless case.  This issue is not merely an artifact of the particular asymptotic form we used for the heat kernel, however. The continuation of R\'enyi entropy to geometric entanglement entropy is dependent on a family of conic metrics. In the limit $m\to 0$, or equivalently, $R\to \infty$, all conic metrics, $g_\delta$, appear to be equivalent near the cone points. This is of course consistent with the notion that, in a CFT, rescaling the metric does not alter the (replica) partition function with which one computes the entropy.  For relevant discussions of geometric entropy in CFTs, we refer the reader to Ref.~\cite{heatkernel,CASINI2010,SOLODUKHIN2008}.

To make contact with the work of Li and Takayangi, we consider a fractional CFT on the sphere
\begin{equation}
	I=\int_{S^d} \phi (-\Delta)^\gamma\phi.
\end{equation}

The geometry that follows from the replica trick is a covering of the sphere ramified at $\Sigma$. The structure near the singular hypersurface is again $C_\delta\times \Sigma$ such that when the deficit angle is $2\pi$, the space is completely smooth. The heat kernel trace can be decomposed as follows~\cite{furasev}

\begin{equation}
\begin{aligned}
\text{Tr}K_\gamma(s) = &\int_{M\backslash \bigcup_i(\Sigma^i\times C_\delta)}d^dx \text{ tr}K_\gamma(x,x,s)\\&\hspace{5mm}+\sum_i\int_{\Sigma^i\times C_\delta}d^dx\text{ tr}K_\gamma(x,x,s),
\end{aligned}
\end{equation}
where $i$ indexes the singular points of the replica space. If at this point, we introduce a correlation length, $\xi$, and take the sub-region $A$ to be a hemisphere, the large-sphere-radius limit implies that one can now asymptotically trace the fractional heat kernel on a space that looks locally like $C_\delta\times \mathbb{R}^{d-2}$. Thus, to leading order, this recovers the computation we have carried out in the massive case. The argument does not hold under $\xi\to \infty$ and hence one has to carry out the appropriate trace for the chosen partitioning of the space. As explained in sec.~\ref{sec:replica}, this cannot be done exactly. Within the heat kernel framework, the obstruction is the fact that no closed analytical expression for the heat kernel trace on the cone exists~\cite{VASSILEVICH2003279}. However, exploiting the conformal invariance, we can rescale the conic metric on $S^1 \times \Sigma $ 
\begin{equation}
dr^2 +  r^2 d\tau ^2 + g_\Sigma\to \left(\frac{dr^2 + g_\Sigma } {r^2} \right)+ d\tau ^2 
\end{equation}
where we recognize the first factor as a conformally compact metric (i.e. asymptotically AdS). Using this metric in the previously outlined replica trick does not result in a conic singularity and hence avoids the problem of requiring a closed form expression for the partition function on the cone while also making the EE insensitive to the choice of $\gamma$. See Ref.~\cite{CASINI2010} for the details of this approach.  

Alternatively, one can continue $N\to1/N$ and trace the heat kernel on the resulting orbifold, $S^d/\mathbb{Z}_N$. The details of the calculation of the EE for the free fractional orbifold theory has been explicated clearly in Ref.~\cite{lt2011} so we do not repeat it here. The EE scaling is (in our notation) 
\begin{equation}
	S\sim \kappa_{d-2}\left(\frac{1}{\epsilon}\right)^{d-2} + \cdots
\end{equation}
thereby leading to an area low of the UV divergence of entanglement entropy despite the fractional Laplacian.

\section{Final remarks}
These calculations highlight also one of the main differences anticipated in the introduction between the two types of theories.   In the massless theory, one can probe the entangling surface only up to a given scale, namely the cutoff scale in a manner independent of $\gamma$ due to the invariance of the conic metric under rescalings. On the other hand, as evident from Eqs. \eqref{mass-ent1} and \eqref{mass-ent2}, in the massive case, one can arbitrarily probe the UV physics. 

To conclude, only B-type theories admit a field redefinition or equivalently a Hilbert-space transformation that exposes the underlying Gaussian nature of the QFT.   When Eq. (\ref{rule}) fails, the theory is truly non-local and the related EE deviates strongly from an area law.   In some cases (C-type theories), we find even a volume law.   Whether or not all deviations\cite{shor2016} from area laws can be understood as a general case of type-C theories is an unanswered question.

We are thankful to Edward Witten for a careful reading of an earlier draft and T. Takayanagi and Luke Yeo for useful conversations and  NSF DMR-1461952 for partial funding of this project.

\appendix*

\section{Fractional heat kernel asymptotics}

Here we want to show the asymptotics of the fractional heat kernel on the 2D cone $C_\delta$. More explicitly we want to justify Eq. \eqref{conicheatexp}. In order to do that we show that there is a (unique) fractional heat kernel $K_\gamma (x,y,s)$ satisfying 
\beq
\left\{ \begin{aligned}&\left(\partial _s - \Delta _x^\gamma\right) K_\gamma(x,y,s)= 0\\&\lim _{s\to 0^+} K_\gamma (x,y,s) = \delta _y(x)\end{aligned}\right.
\eeq
such that as $ s\to 0$
\beq\label{fracheatexp}
 \begin{aligned}K_\gamma(x,x,s)&= s^{-1/\gamma}\left( \sum _k a_k(x) s^k\right) \end{aligned}
\eeq
uniformly in $x$. Then the result follows from the same arguments done for the regular Laplacian on the cone $C_\delta$.
In order to prove the existence of such a heat kernel, one must construct a {\it parametrix} starting from the classical result that the fractional heat kernel in $\mathbb R^d$ is
\beq 
K_{\mathbb{R}^d}= \frac{s}{\left( |x-y| + s^{\frac{1}{2\gamma}}\right) ^{d+2\gamma}}.
\eeq
Next, specializing to $d=2$ and introducing the conic metric, $g$, 
\beq 
K_\gamma=\frac{s}{\left( |x-y|_g(y) + s^{\frac{1}{2\gamma}}\right) ^{2+2\gamma}}
\eeq
where $|x-y|_g$ indicates the distance between $x$ and $y$ with respect to the conic metric $g$ evaluated at the point $y$ (assuming it is not the vertex of the cone), one can show that $K_{\gamma}$ satisfies 
\beq
\left\{ \begin{aligned}&\left(\partial _s- \Delta _x^\gamma\right) K_{\gamma}(x,y,s)= R(s,x,y)\\&\lim _{s\to 0^+} K_\gamma (x,y,s) = \delta _y(x)\end{aligned}\right.
\eeq
The next step is to make $K_{\gamma}$ into an exact solution by summing a convergent series (the
Volterra series). This shows that Eq. \eqref{fracheatexp} holds. Integrating, one gets the desired expansion \eqref{conicheatexp}. In order to achieve that one makes use of the form of the metric $\beta^2 |z|^{2\beta-2} |dz|^2$ (in complex coordinates) with $\delta=2\pi \beta$ and therefore that the metric conical Laplacian is $ \beta^{-2} |z|^{2-2\beta}\Delta$.

%


\end{document}